# Light-induced scattering and energy transfer between orthogonally-polarized waves

**M.A. Ellabban**

Nonlinear Physics Group, Faculty of Physics, University of Vienna, Boltzmanngasse 5, A-1090 Vienna, Austria and Physics Department, Faculty of Science, Tanta University, Tanta 31527, Egypt.

## Abstract

We present a detailed experimental investigation on polarization-isotropic and polarization-anisotropic holographic scattering in lithium niobate crystal doped with iron when recording parasitic gratings with an ordinary polarized pump beam. The kinetics of both types of scattering during the whole process of recording is studied. Holographic scattering is presented as a simple technique to monitor the energy transfer between beams of different polarization. Moreover, the spectral and the angular dependence of the transmitted intensity of the crystal during the reconstruction of the auto-generated parasitic gratings are measured.

# 1. Introduction

The propagation of a laser beam through a photorefractive medium is usually accompanied with unintentional light-induced scattering, so-called holographic scattering. It originates from recording parasitic gratings as a result of the interference between the incident beam and its own scattered beams generated within the photorefractive medium. On one hand, the presence of parasitic gratings is considered to be a major obstacle for the applicability of thick photorefractive media in data storage and image processing applications as it corrupts the quality of any optical system: It transfers almost all the light from the initial beam to the scattering directions and adds strong noise to the reconstructed images. Therefore, eliminating or at least suppressing the effect of holographic scattering becomes a main goal in order to promote photorefractive crystals for industrial applications [1]. On the other hand holographic scattering has been advantageously utilized for a wide variety of applications such as optical limiters [2,3], color or coherent-to-incoherent converters [4,5], novelty filters [6,7], vibration object monitoring [8], logic operations [9,10], hologram multiplexing techniques that use random wave generated by beam fanning instead of external diffusers [11] and material characterization [12-16].

In the standard photorefractive media (electro-optic media) the anisotropy of the physical properties makes the currents generated by the incident light within the medium being dependent on the intensity, the polarization, and the orientation of the incident wave with respect to the crystallographic axes. Polarization-isotropic holographic scattering that is induced within a photorefractive crystal by an extraordinary polarized pump beam has been extensively studied (see, for example, [13] and [15]). So far, only few investigations have been carried out on parasitic gratings recorded with ordinarily polarization beam. However, in the most common geometry for two-wave mixing experiments with photorefractive crystals an elementary grating is recorded with two ordinary polarized waves.

When a beam with ordinary polarization is directed onto a model photorefractive crystal like lithium niobate doped with iron at first a strong polarization-isotropic scattering distributed along the c-axis appears. Then a symmetrical scattering with an extraordinary polarization grows up with recording time. It is distributed perpendicular to the c-axis and is accompanied by a reduction in the intensity of scattered light with ordinary polarization. If the light polarization is preserved in the scattering process, the scattering is said simply to be isotropic otherwise anisotropic. The latter was first reported [17] and analyzed by Avakyan et al. [18]. The analysis is a two-wave grating efficiency calculation that is based on the bulk photovoltaic effect. Later on Wilson et al made the analysis using the bulk photovoltaic model to derive coupled wave equations either for arbitrary-direction two-beam coupling or multiple-beam coupling including the effect of beam diameter [19]. This anisotropic scattering is due to the recording of parasitic gratings by the excitation of a spatially oscillating photovoltaic current that is generated by the interaction of orthogonally polarized beams [18]. Odoulov already built up a dynamic grating in a lithium niobate crystal doped with iron with two orthogonally polarized beams which is an experimental proof for the existence of spatially oscillating currents in crystals which have non-diagonal components of the photovoltaic tensor [20]. Thus, anisotropic scattering can be used to estimate the non-diagonal photovoltaic tensor components in different electro-optic crystals [20,21] which can not be done with the usual electrical methods. Odoulov et al. observed also anisotropic scattering in $LiTaO_3$:Cu from an extraordinary polarized pump beam with the rotation of the plane of polarization from extraordinary to ordinary [21]. Moreover, the reconstruction process in elementary holography applications can be performed at a replay wavelength or angle which differs from the recording one, e.g. angular [22] and wavelength multiplexing [23] techniques. Thus, it is important to study the angular and spectral dependence of the strength of the auto-generated parasitic gratings. Additionally, the ordinary-extraordinary scattering is investigated only in the steady state not at different times of recording [17-19]. Furthermore, the ordinary-ordinary scattering that occurs before the ordinary-extraordinary scattering dominates is not analyzed at all so far [17-19]. For all above mentioned reasons and the deficient in information about the energy transfer from the ordinary to the extraordinary scattered light, it is interesting to investigate the recording process of the parasitic gratings by an ordinary recording beam incident on a lithium niobate crystal doped with iron and then reconstructing these gratings at different conditions.

In this work we study the build up or decay of different types of scattering during recording of parasitic gratings with an ordinary polarized pump beam. Then, we measure the angular and the spectral dependence of the transmitted intensity during readout of the written gratings. The parasitic gratings are recorded and reconstructed using a beam parameters that are usually applied with recording and reconstructing the elementary gratings, plane waves and low or moderate intensity.

## 2. Experimental

An expanded beam of an Ar-ion laser is used to record and reconstruct parasitic gratings at a wavelength of $\lambda p=488$ nm and beam diameter 4 mm. The investigated sample is an a-cut lithium niobate crystal doped with iron at a doping level of $Fe^{2+}$ of $7.2 \times 10^{23}$ m$^{-3}$ and dimensions of 17.90 (a) x 2.55 (b) x 9.60 (c) mm$^3$. The pump beam is ordinarily polarized and incident perpendicular to the crystal surface with an intensity of 0.14 W/cm$^2$. The polarization and the intensity of the pump beam were adjusted with a combination of $\lambda/2$ plate and a polarizer. The scattering patterns of isotropic and anisotropic scattering are projected onto a screen placed 8.5 cm behind the crystal and registered by a CCD camera during the recording time.

A quantitative measurement of light-induced scattering can be made via measuring the transmitted intensity of the crystal. This technique is efficient and unique if it is not easy to predict where the scattering will occur. Reconstruction of parasitic holograms was performed at different wavelengths, $\lambda r$, using a probe beam with a low intensity that is four orders of magnitude less than the pump intensity. The angular dependence of transmitted intensity was measured when the crystal was rotated around an axis perpendicular to the *c*-axis, $\phi$-rotation. The sample holder was fixed on an accurately controlled rotation stage ($\pm 0.001°$).

## 3. Results

As a first step we registered the far-field scattering pattern, isotropic and anisotropic scattering, at different recording times. A polarization sheet is used as an analyzer of the scattered light. Some examples are shown in Fig. 1. At first the pump beam with ordinary polarization distorts asymmetrically along the c-axis (more clear in Figs. 1(b) and (c)). The incident and scattered light show identical polarization and the scattering lobes are inclined to the c-axis with approximately 10°. Then after attaining a maximum intensity, a symmetrical scattering with extraordinary polarization perpendicular to the polar axis grows up (Figs. 1(h)-(j)) with time and is accompanied by a reduction in the intensity of the scattered light of ordinary polarization.

Isotropic scattering decays at large angles with time while the anisotropic scattering is growing. At the end of recording there is still a strong small angle isotropic scattering in comparison to the anisotropic if the exposure time is taken into account. In order to find a quantitative measure for the isotropic and anisotropic scattered intensity and its distribution along different directions, we analyzed the two-dimensional scattering patterns displayed in Fig. 1 and others patterns which are not shown in Fig. (1) along $k_z$ at $k_x = 0$ and along $k_x$ at $k_z = 0$, indicated by the thin and thick white lines, respectively. The analysis takes into account the exposure time of each photograph. Figs. 1(f)-(j) show that there is nearly no scattering of extraordinary polarization at the beginning of the recording process. It is clearly shown that the extraordinary scattered intensity grows even along the direction of the c-axis. Then it slightly grows nearly in all directions, Fig. 1(g) and (h), and finally the scattered light is amplified mainly in the direction perpendicular to the optic axis whereas it is depleted completely in other directions. When analyzing the ordinary-extraordinary scattering it is usually assumed that there is a small fraction of the ordinary light at the input of the crystal that is converted to a uniform angular distribution of extraordinary light [19]. This assumption is proved in Figs. 1(g) and (h).

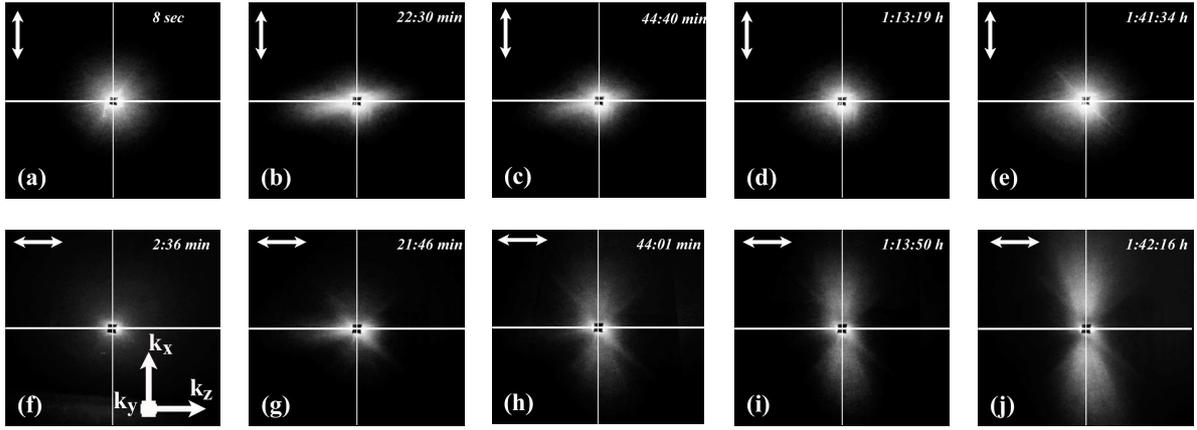

**Figure 1.** Far-field scattering patterns observed at different recording times of an ordinary polarized pump beam. The central black spot corresponds to the transmitted beam. The arrows indicate the polarization direction of the polarizer sheet placed behind the crystal. The exposure time for the photographs f-j is nearly one order of magnitude larger than that for the photographs a-e. The +c-axis is parallel to $k_z$.

Fig. 2 and Fig. 3 show the distribution of the scattered intensity in different directions which is obtained from the analysis of the scattering patterns at various recording times. The sudden drop of the intensity at small angles around zero is due to the black block placed on the screen in the direction of the transmitted beam. As shown in Fig. 2, the isotropic scattered intensity at first significantly increases until a recording time of 22.5 minutes, then it slightly decreases until the end of recording, but there remains till a strong isotropic scattering (Fig. 2, stars). The ratio of the integrated scattered intensity in the scattering pattern along $k_z$ at the end of recording (1.69 hours) and time corresponds to the maximum scattered intensity (22.5 minutes) is about 0.5. One can see also in Fig. 2 that the isotropic scattering is asymmetrically distributed along both directions of the c-axis. There is more scattered intensity in the direction of the -c-axis. The asymmetry and the scattering angles corresponding to maximum scattered intensities change in a similar way to the scattered intensity. There is till some asymmetry of the scattered intensities at the end of recording. The analysis along $k_x$ does not reveal any significant scattering outside the region of the transmitted beam. This can be seen also in see Figs. 1(a)-(e) which do not show any scattering lobes along $k_x$ in contrast to those strong lobes generated along $k_z$. Let us now switch to the analysis of the anisotropic scattering along $k_z$ and $k_x$. Fig. 3(a) shows that the scattered intensity grows slightly and symmetrically with the recording time along $k_x$ until saturation started at recording time of 1.54 hours. Note that the scattered intensity is much smaller than that for isotropic scattering along $k_z$, the ratio between the maximum integrated scattered intensities along $k_x$ for anisotropic scattering and $k_z$ direction (isotropic scattering) is about 0.08.

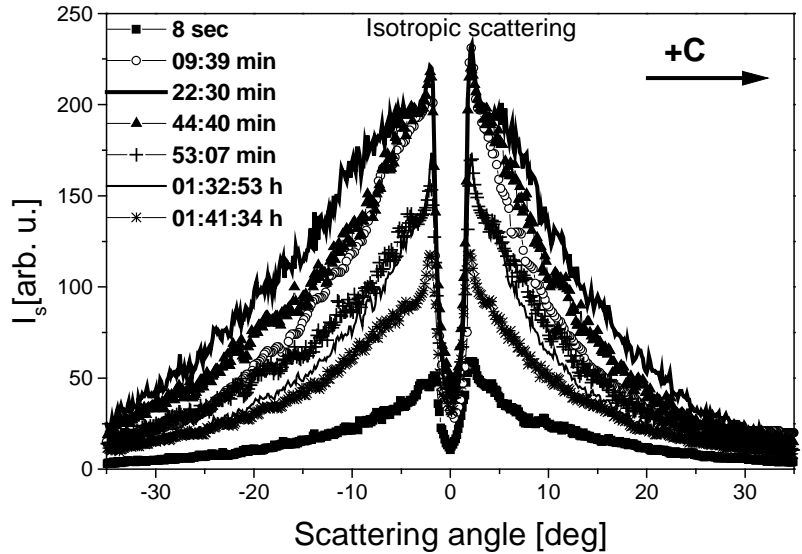

**Figure 2.** Scattered intensity distribution along $k_z$ at $k_x = 0$ (thick white lines in Fig. 1) for isotropic anisotropic scattering.

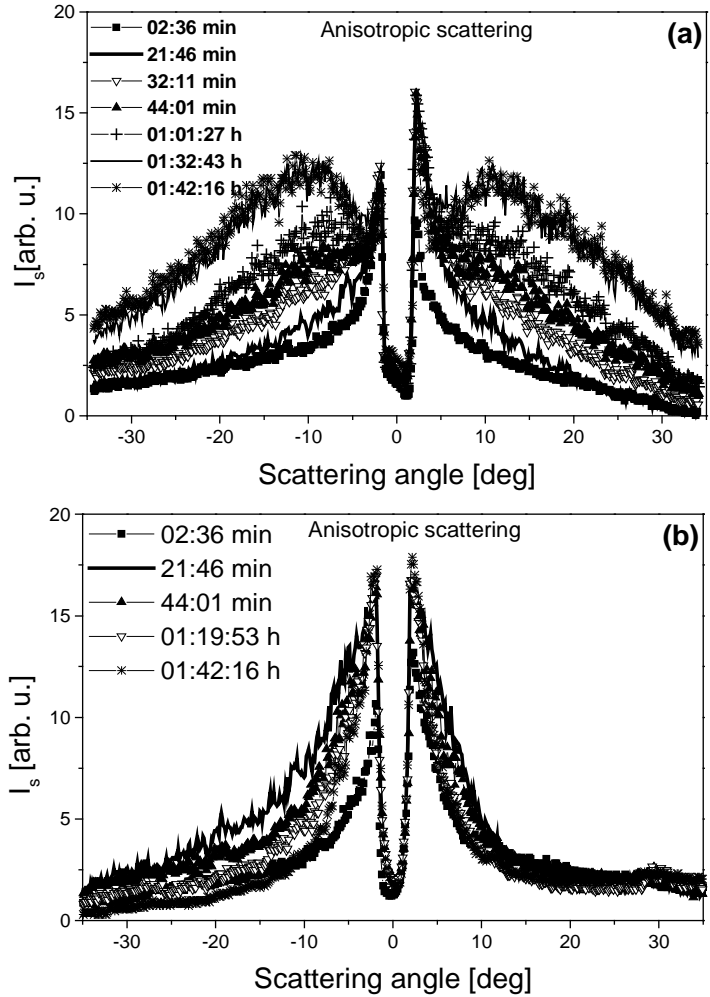

**Figure 3.** Scattered intensity distribution of anisotropic scattering along (a) $k_x$ at $k_z = 0$ (thin white lines in Fig. 1) and $k_z$ at $k_x = 0$ (thick white lines in Fig. 1).

However, the intensity of anisotropic scattering is distributed along larger than that for isotropic scattering. Fig. 3(b) shows that there is anisotropy scattering along the c-axis. There was nearly no scattering at recording time 2.6 minutes, but it significantly increases until recording time 21.75 minutes and then slightly decreases until it nearly vanishes at the end of recording.

In a second step we measured the angular dependence of transmitted intensity, normalized to transmitted intensity at large angles, at different reconstruction wavelengths and the polarization of recording was kept the same. These results are shown in Fig.4 (a). The lowest transmitted intensity occurs when the reconstruction conditions is similar to that of recording ($\lambda_r=\lambda p$) and the transmittance curve exhibit only one minimum even at the end of recording we observed anisotropic scattering. This is another confirmation after that obtained from analyzing the scattering pattern that the dominant scattering is isotropic. The transmittance curves exhibit two minima for wavelengths lower than the writing wavelength and the depths of the two minima are not the same. That asymmetry reflects the small asymmetry in the scattering pattern along the *c*-axis. For $\lambda r>\lambda p$, the transmittance curve has one minimum that is not symmetric with respect the Bragg angle. However, this minimum can be an envelope of a double minimum structure which cannot be resolved [13]. Figure 4(b) shows that detuning the readout angle by approximately 0.5° restores the initial transmittance for $\lambda_r=\lambda_p$. This angular selectivity is higher than in the case of parasitic gratings written with extraordinary polarized beam and the full width at half maximum of the transmittance curve for ordinary-polarization written gratings is 0.4 from that written by extraordinary polarization.

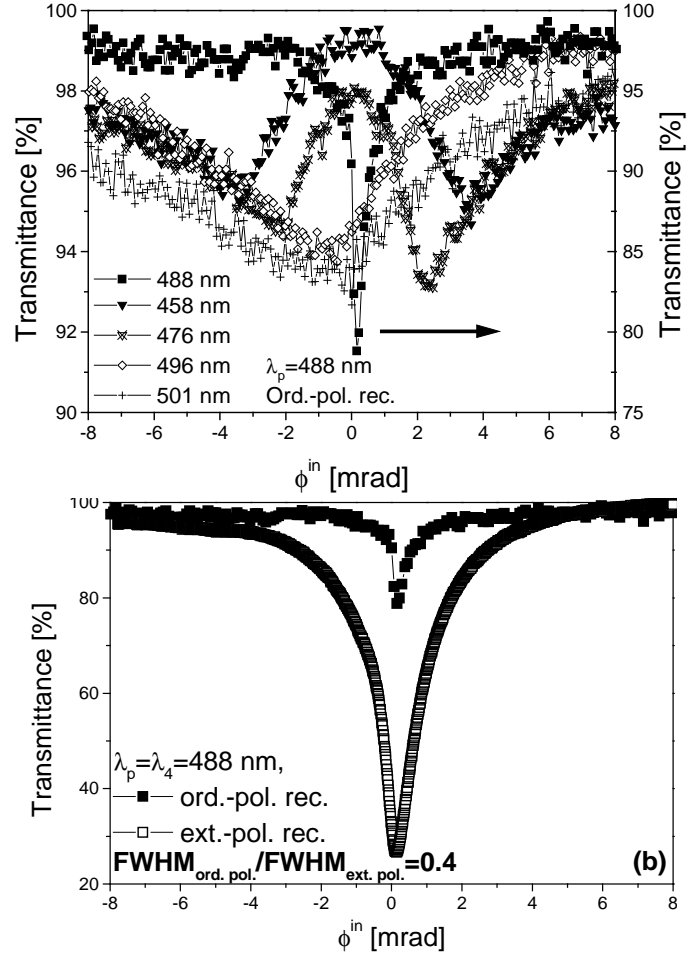

**Figure 4.** Normalized transmittance as a function of the readout angle at (a) different reconstruction wavelengths for parasitic gratings written with ordinary-polarization recording beam and (b) at $\lambda_r=\lambda p$ for gratings recording with ordinary or extraordinary-polarized pump beams. The angle is given within the medium and recording was done at $\phi=0$.

## 4. Discussion

At first we will discuss briefly the isotropic and anisotropic scattering for parasitic gratings recorded with an ordinary polarized beam. Then we will discuss the characteristic curves of reconstructing these written gratings at different conditions.

The electric field of the incident beam produces a current whose magnitude and direction depend on the intensity, the orientation and the polarization of the light wave. It is given by

$$j_i^{ph} = \beta_{ijk} E_j E_k^*, \qquad (1)$$

where $j_i^{ph}$ is the photovoltaic current density, $\beta_{ijk}$ is the photovoltaic tensor and $E_j$ and $E_k^*$ are the amplitudes of the electric field component of one of the interfering light waves and the complex conjugate of the other. Interference between the pump beam and one of its scattered waves which has the same polarization spatially modulates the current according to the intensity modulation and creates a space charge field

$$E_i^{sc} = j_k^{ph} \sigma_{ki}^{-1}, \qquad (2)$$

where $\sigma$ is the photoconductivity. The space charge field in turn modulates the refractive index via the linear electro-optic effect:

$$\Delta(\varepsilon_{ij}) = r_{ijk} E_k^{sc}, \qquad (3)$$

where $\varepsilon=(n)^2$ is the dielectric permittivity tensor, n is the refractive index, and $r_{ijk}$ is the linear electro-optic tensor.

For the geometry we used in recording the parasitic gratings with an ordinary polarized beam, the isotropic scattering has two components [12]. One is due to $r_{13}\beta_{31}$ and the other to $r_{22}\beta_{22}$, both are linear photovoltaic tensor elements. The component $r_{22}\beta_{22}$ generates grating vectors perpendicular to the optic axis whereas $r_{13}\beta_{31}$ produces grating vectors along the polar axis. The scattering pattern shown in Fig. 1 indicates that there is no significant isotropic scattering perpendicular to the c-axis. This is in accordance with the small ratio of $r_{22}\beta_{22}/r_{13}\beta_{31}$ which is about 0.03 as calculated from values available in literature [24]. The scattering occurs mainly along the polar axis with a slight inclination, nearly 10°. This inclination is due to the transverse field induced by $\beta_{22}$ [25]. The asymmetry of the scattering pattern for isotropic scattering shown in Fig.1 reflects the relative contributions of the photovoltaic effect and the diffusion to the charge transport processes during writing the gratings of the isotropic scattering. Diffusion leads to a steady state unidirectional energy transfer from the pump beam to the scattered beams along the – c-axis for our crystal where the electro-optic coefficient is positive and the charge carriers are electrons. This is not the case for the photovoltaic effect which results in a transient energy transfer between the pump beam and the scattered beams symmetrically in both directions of the c-axis. The transient energy transfer is due to the interaction between beams of dissimilar intensities, the pump and the scattered beams. Also, seed scattering amplification due to local response may be important if there is low frequency noise, possibly due to light intensity and/or refractive-index fluctuations [25].

When the pump wave with ordinary polarization interacts with an extraordinary polarized scattered beam, the intensity distribution within the crystal is uniform, but the polarization state is modulated. Thus, spatially oscillating currents in directions perpendicular to the polar axis are produced via the non-diagonal photovoltaic component $\beta_{51}$. The magnitude and direction of this current depends on polarization of the light wave within the crystal. These currents generates a space charge field that modulates the refractive index via the linear electro-optic effect, i.e. a weak grating whose wave vector is perpendicular to the c-axis is recorded. This grating amplifies the scattered wave as a result of self diffraction of the incident beam by a grating that is displaced with respect to the positions of linear polarization due to the imaginary antisymmetric part of the photovoltaic tensor element, $\beta^c_{51}$. This element gives a non-local contribution to the current density responsible for holographic amplification of anisotropic scattering. Thus, energy exchange between the ordinary pump wave and the extraordinary scattered wave is allowed, the oscillating current increases and in turn the grating amplitude becomes stronger. This process continues till a steady state is reached. Fig. 5 sketch simply recording one of the possible gratings for isotropic and anisotropic scattering in the $k_x$-$k_y$ plane in a negative birefringent crystal. We want to draw the attention here to the point, that the homogeneity of the intensity through the crystal does not permit contribution from diffusion to the charge carriers processes and the only contribution is the photovoltaic effect which produces a symmetric scattering.

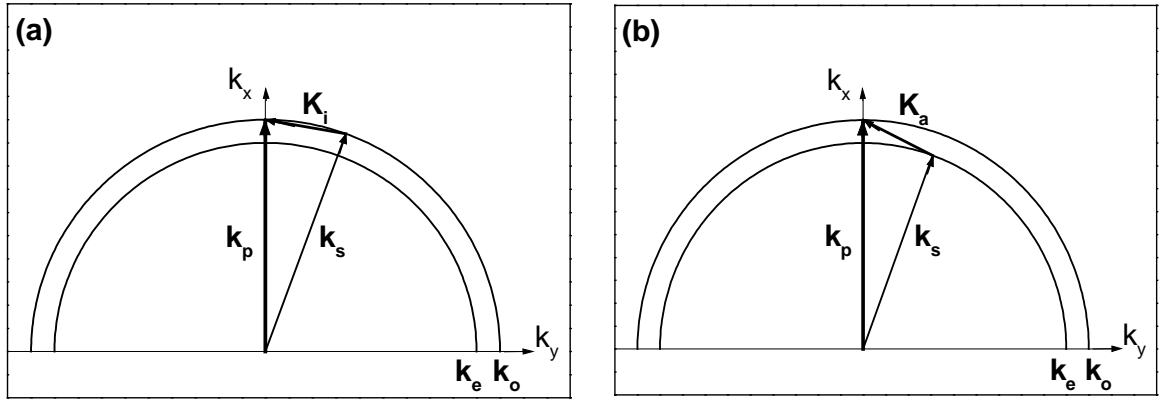

**Figure 5**. Vector diagram in the reciprocal space for (a) isotropic and (b) anisotropic scattering in the $k_x$-$k_y$ plane, $\mathbf{k_p}$, $\mathbf{k_s}$, $\mathbf{K_i}$ and $\mathbf{K_a}$ are the pump, scattered wave, the isotropic and the anisotropic gratings, respectively.

The gain factor that governs the transfer of energy from the pump to the scattered wave is given by

$$\Gamma_{o,e} = \frac{\pm 2\pi n^3 r_{51} \beta^c_{51}(\mathbf{K_a e_{o,e}})}{\lambda_\mathbf{p} k_{ph} \cos(\theta_{o,e})}, \quad (4)$$

$k_{ph}$ is the specific photoconductivity, $\mathbf{K_a}=\mathbf{K_a}/|\mathbf{K_a}|$ and $\mathbf{e_{o,e}}$ is the ordinary or extraordinary polarization unit vector and $\theta_{o,e}$ is the angle between the pump and scattered beam with ordinary or extraordinary polarization [25]. The sign of the product $r_{51}\beta^c_{51}$ uniquely defines the direction of intensity transfer between waves with different polarizations. Since the extraordinary polarized mode is amplified at the expense of the ordinary polarized mode in LiNbO3:Fe, then $\beta^c_{51}$ must be negative whereas energy transfers from the extraordinary to ordinary in LiTaO$_3$:Cu so that in this case $\beta^c_{51} > 0$ [21]. The gain described in the previous equation does not depend on the pump intensity, but it should be noted that the photovoltaic current density itself is proportional to the light intensity (Eq. (1)).

We did not observe a complete ordinary-to extraordinary conversion in our sample or another sample with nearly the same content of $Fe^{2+}$ but with a thickness 1.3 mm. This is in accordance with the findings of Wilson et al. that getting a complete conversion needs a long interaction length and high doping levels [19]. The energy transfer depends on the photovoltaic tensor element $\beta^c_{51}$ and the latter is proportional to the concentration of filled traps and depends on the photon energy [26] (Eq. (4)). However, we can estimate the value of $\beta^c_{51}$ for our crystal from the ratio of the anisotropic to isotropic, at maximum. The ratio should be given by $r_{51}\beta^c_{51}/r_{13}\beta_{31}$, $\beta^c_{51}$ should be 1.4x10-9A/W if the ratio is 0.08 and using the known values of the other parameters from literature [24].

Now we will discuss the results of reconstructing parasitic gratings at different conditions. Figs. 1-3 point out that the scattering is mainly isotropic. This was confirmed also in Fig. 4(a) where the maximum scattered intensity occurs at $\lambda_r=\lambda_p$. Also, the transmittance curves have a two minima structure for readout wavelengths that differs from the writing one. All these features were explained by the Ewald sphere construction model for the case of isotropic scattering built up from recording parasitic gratings with an extraordinary polarized pump beam [13]. In both cases, the scattering is isotropic and the rotation is done around an axis that is perpendicular to the c-axis, but there two differences: the polarization of the recording beam is different (ordinary in our work and extraordinary in Ref. [13]) and the scattering in our work is nearly three dimensional whereas it is two dimensional in Ref. [13]. These differences do not affect the simulated transmittance qualitatively. Thus, our results can be explained qualitatively in similar to the work of Ref. [13] by Ewald sphere construction model. The model is presented in detail in Ref. [27].

Fig. 4 (b) shows that the ratio of the scattered intensity from parasitic recorded by ordinary polarized beam to that recorded by extraordinary polarized beam at the Bragg angle is 0.29 which is exactly the ratio $r_{13}\beta_{31}/r_{33}\beta_{33}$ as calculated from literature [24]. We like to pay the attention to the point that holographic scattering technique is an accurate and simple technique for material characterization. Fig. 4. points out that the light-induced scattering is suppressed significantly at the Bragg angle for gratings written by ordinary polarization. It is nearly four times less than scattering produced by gratings recorded with extraordinary polarized beam. The suppression of light-induced scattering is an essential requirement in optical data storage and image processing applications. Also, Fig. 4(a) indicates that the maximum loss obtained at the reconstruction of parasitic gratings at different wavelengths for readout angles is less than 0.07. On the other hand, the extraordinary recording of parasitic gratings is useful if we want to benefit from light-induced scattering in different application such as material characterization. For the letter holographic scattering technique is superior in comparison to other holographic techniques: it is simpler and does require high demand of mechanical stability.

## 5. Conclusion
In summary, we present a detailed study of the parasitic gratings recorded with an ordinary polarized beam in a lithium niobate crystal doped with iron. We conduct the first study on energy transfer from the ordinary to the extraordinary scattered waves. We already detected a small fraction of the ordinary light that is converted to a uniform angular distribution of extraordinary scattered light which is important as a seed for further conversion. Light-induced scattering is a useful, accurate and simple technique to monitor the energy conversion from one beam to another and also to get information about the material parameters, particularly the non-diagonal photovoltaic tensor components which can not be determined using the standard electrical methods. Recording with ordinary polarized beams suppresses the effect of the parasitic gratings at different reconstruction conditions whereas writing with extraordinary polarized beams produces a strong effect when reconstructing parasitic gratings which can be useful for material characterization.

## Acknowledgment

The financial support of the Austrian Science Fund FWF(P-18988) is acknowledged. The author is grateful to Prof. M. Fally for his critical reading of the manuscript.